\documentclass[aps,prev,twocolumn,preprintnumbers,floatfix,nofootinbib]{revtex4-1}

\usepackage{graphicx}
\usepackage[english]{babel}
\usepackage{bm}
\usepackage{times}
\usepackage{slashed}
\usepackage{mathtools}
\usepackage{color}
\usepackage{epsfig}
\usepackage{dcolumn}
\usepackage{textcomp}
\usepackage{color}

\def\hmath$#1${\texorpdfstring{{\rmfamily\textit{#1}}}{#1}}
\usepackage{hyperref}
\usepackage{amsfonts}
\usepackage{threeparttable}

\begin{document} 

\preprint{IIPDM-2019}

\title{ New Physics Probes: Atomic Parity Violation, Polarized Electron Scattering and Neutrino-Nucleus Coherent Scattering}

\author{Giorgio Arcadi$^{1,2}$} \author{Manfred Lindner$^{3}$}
\author{Jessica Martins$^4$} \author{Farinaldo S. Queiroz$^5$}

\affiliation{$^1$ Dipartimento di Matematica e Fisica, Universit\`a di Roma 3, Via della Vasca Navale 84, 00146, Roma, Italy}

\affiliation{$^2$INFN Sezione Roma 3}
\affiliation{$^3$Max-Planck-Institut f\:ur Kernphysik (MPIK), Saupfercheckweg 1, 69117 Heidelberg,
Germany}
\affiliation{$^4$Instituto de F\'isica Te\'orica,
Universidade Estadual Paulista, S\~ao Paulo, Brazil}
\affiliation{$^5$International Institute of Physics, Universidade Federal do Rio Grande do Norte, Campus Universit\'ario, Lagoa Nova, Natal-RN 59078-970, Brazil\\
}

\email{giorgio.arcadi@uniroma3.it}
\email{manfred.lindner@mpi-hd.mpg.de}
\email{jessica.martins@unesp.br}
\email{farinaldo.queiroz@iip.ufrn.br}

\begin{abstract}
Atomic Parity Violation (APV) is usually quantified in terms of the weak nuclear charge $Q_W$ of a nucleus, which depends on the coupling strength between the atomic electrons and quarks. In this work, we review the importance of APV to probing new physics using effective field theory. Furthermore, using $SU(2)$ invariance, we correlate our findings with those from neutrino-nucleus coherent scattering. Moreover, we investigate signs of parity violation in polarized electron scattering and show how precise measurements on the Weinberg angle, $\sin \theta_W$, will give rise to competitive bounds on light mediators over a wide range of masses and interactions strength. Lastly, apply our bounds to several models namely,  Dark Z, Two Higgs Doublet Model-$U(1)_X$ and 3-3-1, considering both light and heavy mediator regimes.
\end{abstract}

\maketitle
\flushbottom

\section{Introduction}
\label{introduction}

For decades it was assumed that the laws of nature preserved parity, but the seminal paper of
Lee and Yang in 1956 gave rise to a different perspective \cite{Lee:1956qn}. This was indeed confirmed in 1957 in the realm of weak interactions, via the beta decay in Cobalt \cite{Wu:1957my} and muon decay \cite{Garwin:1957hc}. In 1959 the possibility of observing parity violation in atomic physics and electron scattering was contemplated \cite{bib:Zeldovich:59} and further investigated in the late 70's \cite{Prescott:1978tm,Barkov:1978fb,Bouchiat:1982um}. Interestingly these ideas preceded the theory of electroweak interactions. The following decades were populated by experiments that aimed at probing parity violation \cite{Erler:2014fqa}. For interesting reviews on APV see  \cite{Bouchiat:1997mj,RamseyMusolf:1999qk,Rosner:1999cy,Rosner:2001ck,Ginges:2003qt,Derevianko:2006sf}.    

Our current understanding of parity and APV has greatly improved. Given the experimental precision acquired over the years, we may now test new physics models that feature parity violation interactions. The main objects of study in this regard are the Weak Charge of the Cesium (Cs) and polarized electron scattering. 

Concerning the first observable, the weak charge of a nucleus, $Q_W$, is an analogous of the electromagnetic charge, where the Z boson is the key player of the atomic electron and nucleus interactions instead. $Q_W$ is the sum of the weak charges of all constituents of the atomic nucleus, $Q_W = (2Z+N) Q_W (u) + (Z+2N) Q_W (d)$, where $Q_W(u,d)$ accounts for the Z interactions with up and down quarks and it depends on the Weinberg angle, $\sin^2\theta_W$. In order to understand how the Z boson can affect atomic transitions and $Q_W$ is extracted from experiments, one needs to perform precise atomic physics calculations and measure the left-right asymmetry $A_{LR}$, which is naively estimated to be $A_{LR} \simeq \alpha^2 m_e^2/m_Z^2 \sim 10^{-15}$. Despite atoms with high atomic number be better suited for experimental observation of APV \cite{Guena:2005uj} because of the enhancement in $A_{LR}$ by orders of magnitude,  such high atomic number makes the theoretical determination of $A_{LR}$ extremely challenging. For this reason, Cesium has become a popular target because it offers a good compromise between high atomic number necessary to have sizable effects and relatively simple atomic
structure required to make precise atomic calculations. The Standard Model prediction for the weak charge of Cesium reads  $Q_W^{th}=-73.16$ \cite{Dzuba:2012kx}, which is in agreement with current measurements using the stable isotope $^{133}_{55}Cs$ \cite{Porsev:2010de}. Therefore, one can constrain new physics contributions using the precise measurements of the Cs weak charge. 

The other parity violation observable, polarized electron scattering, also constitutes an important laboratory to new physics searches. Again the left-right asymmetry is the key observable. For process of the type $e_{L,R} N \rightarrow e X$, the left-right asymmetry is $A/Q^2 = a_1 +a_2f(y)$, where $a_{1,2}$ account for the vector-axial coupling between the electron and quarks which depend on $\sin^2\theta_W$, $y$ the fraction energy transfer from the electron to the hadrons, and $Q$ the moment transfer. Thus, a measurement of A translates into a measurement of $\sin^2\theta_W$ at a given momentum $Q$. It is well-known that photon exchange diagrams conserve parity but processes mediated by the Z do not, since the latter does not interact with left-handed and right-handed fermions in the same way. In a similar vein, eventual additional massive vector bosons from new physics models might also contribute to the left-right asymmetry. Therefore, if the measurement of $A$, in other words, $\sin^2\theta_W$, agrees with the Standard Model prediction one can use this information to constrain new physics effects that induce parity violation and hence contribute to the left-right asymmetry. We can parametrize the new physics contributions to $A$ by a shift on $\sin^2\theta_W$ and consequently constrain new physics effects \cite{Davoudiasl:2012qa}. 

The aforementioned observables depend on the interactions between electrons and quarks. Hence they lead to relevant bounds on the corresponding couplings.  That has been the whole story up to now, but with the observation of neutrino-nucleus coherent scattering new information came into light. Strictly speaking, neutrino-nucleus coherent scattering and parity violation probes are sensitive to different interactions, the former between electron and quarks, the latter between neutrinos and quarks. Nevertheless, using $SU(2)$ invariance, one can potentially correlate the signal in neutrino-nucleus coherent scattering to the one appearing in parity violation observables. In other words, they are complementary to one another. We will carry out this complementarity study using effective field theory and later concentrate on vector mediators. We highlight that our work adds to the previous ones done in the literature because instead of focusing on one observable we explore the complementarity between neutrino-nucleus coherent scattering, APV and polarized electron scattering.  Moreover, we apply our findings to existing models in the literature.\\

In {\it Section 2}, we will review the theoretical aspects of parity violation; in {\it Section 3} we discuss APV; in {\it Section 4} we address the complimentary aspects with neutrino-nucleus coherent scattering using effective field theory; in {\it Section 5} we study polarized electron scattering in terms of light mediators and put into perspective with neutrino-nucleus coherent scattering. Lastly in {\it Section 6} we discuss our bounds using concrete models proposed in the literature.

\section{Parity Violation} 

\begin{figure}[!t]
\centering
\includegraphics[width=\columnwidth]{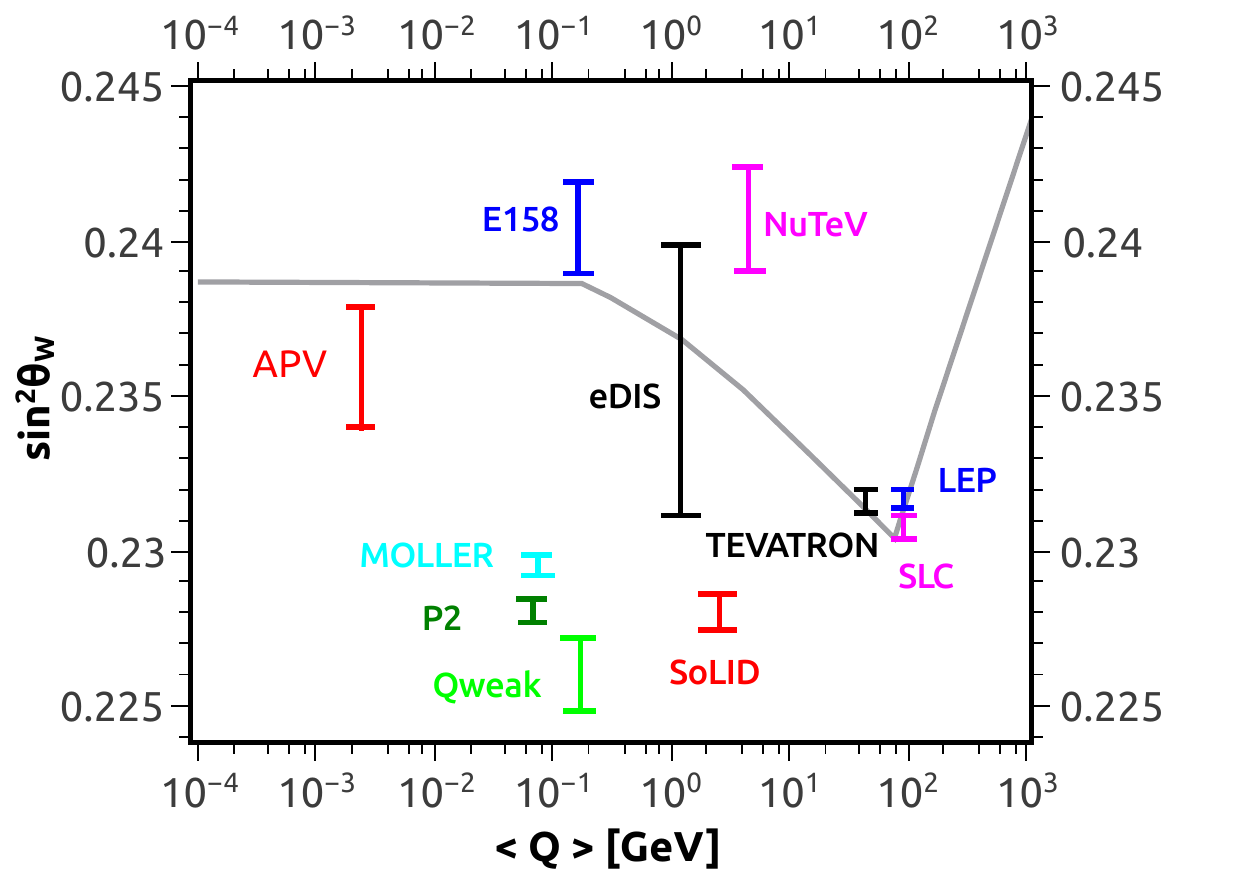}
    \caption{The gray curve reflects the scale dependence of the weak mixing angle. We also show the existing and upcoming measurements \cite{Tanabashi:2018oca}.}
    \label{fig:sintheta}
\end{figure}

We start reviewing how parity can be treated in terms of four-fermion interactions. The bilinear terms that respect Lorentz invariance are, 
\begin{equation}
\mathcal{L}_{\text{int}} = \bar{\Psi}\mathcal{O}\Psi
\end{equation}where,
\begin{equation}
\mathcal{O} = 1,\gamma^{5},\gamma^{\mu},\gamma^{\mu}\gamma^{5}.
\end{equation}

Some combinations respect parity, and the term $\bar{\Psi}\gamma^{5}\Psi\bar{\Psi}\Psi$ is non-hermitian. Therefore we are interested in the terms that involve vector and vector-axial interactions as follows, 

\begin{equation}
\mathcal{L}_{\text{eff}} = \bar{\Psi_a}\gamma^{\mu}\gamma^{5}\Psi_a\bar{\Psi_b}\gamma_{\mu}\Psi_b + \bar{\Psi_a}\gamma^{\mu}\Psi_a\bar{\Psi_b}\gamma_{\mu}\gamma^{5}\Psi_b.
\end{equation}

That said, we shall now understand how parity violation can be described in the neutral current of the Standard Model. In the Standard Model the neutral current reads,

\begin{eqnarray}
\mathcal{L}_{\text{eff}} & =&\bar{e}\gamma^{\mu}\left[\left(\frac{1-\gamma^{5}}{2}\right)T_{3e} - Q_{e}\sin^{2}\theta_{W}\right]e\times\\ &&\bar{q}\gamma^\mu\left[\left(\frac{1-\gamma^{5}}{2}\right)T_{3q} - Q_{q}\sin^{2}\theta_{W}\right]q.
\end{eqnarray}

Separating the terms that induce APV we get,

\begin{eqnarray}
\mathcal{L}_{\text{eff}} =& \frac{1}{2}\bar{e}\gamma^{\mu}T_{3e}e\left(-\frac{1}{2}\bar{q}\gamma_{\mu}\gamma^{5}T_{3q}q\right) - \frac{1}{2}\bar{e}\gamma^{\mu}\gamma^{5}T_{3e}e\left(\frac{1}{2}\bar{q}\gamma_{\mu}T_{3q}q\right) - \nonumber \\
& \frac{1}{2}\bar{e}\gamma^{\mu}\gamma^{5}T_{3e}e(-\bar{q}\gamma_{\mu}Q_{q}\sin^{2}\theta_{W}q)+ \nonumber \\
&\frac{1}{2}\bar{e}\gamma^{\mu}Q_{e}\sin^{2}\theta_{W}e\bar{q}\gamma_{\mu}\gamma^{5}T_{3q}q,
\end{eqnarray}
which simplifies to,

\begin{eqnarray} \mathcal{L}_{\text{eff}}&=&-\frac{1}{2}\bar{e}\gamma^{\mu}e\left(\frac{T_{3e}}{2} - Q_{e}\sin^{2}\theta_{W}\right)\bar{q}\gamma_{\mu}\gamma^{5}qT_{3q} - \nonumber \\
&& \frac{1}{2}\bar{e}\gamma^{\mu}\gamma^{5}eT_{3e}\left(\frac{T_{3q}}{2} - Q_{q}\sin^{2}\theta_{W}\right)\bar{q}\gamma_{\mu}q.\nonumber\\
\end{eqnarray}

Taking $T_{3e}=-1/2,\,Q_{e} = -1,\, T_{3u}=1/2,\, Q_{u} = 2/3,\, T_{3d}=-1/2 ,\, Q_{d}=-1/3$, we find,

\begin{eqnarray}
\mathcal{L}_{\text{eff}} &= &-\frac{1}{2}\bar{e}\gamma^{\mu}e\left(-\frac{1}{4} +\sin^{2}\theta_{W}\right)\bar{u}\gamma_{\mu}\gamma^{5}u\left(\frac{1}{2}\right)\nonumber\\
&& - \frac{1}{2}\bar{e}\gamma^{\mu}\gamma^{5}e\left(-\frac{1}{2}\right)\left(\frac{1}{4} - \frac{2}{3}\sin^{2}\theta_{W}\right)\bar{u}\gamma_{\mu}u \nonumber\\
&& -\frac{1}{2}\bar{e}\gamma^{\mu}e\left(-\frac{1}{4} +\sin^{2}\theta_{W}\right)\bar{d}\gamma_{\mu}\gamma^{5}d\left(-\frac{1}{2}\right)\nonumber\\ &&-\frac{1}{2}\bar{e}\gamma^{\mu}\gamma^{5}e \left(-\frac{1}{2}\right)\left(-\frac{1}{4} + \frac{1}{3}\sin^{2}\theta_{W}\right)\bar{d}\gamma_{\mu}d. 
\end{eqnarray}

Knowing that $\sin^2\theta_W\simeq1/4$, the first and third term are negligible compared to the others, and we are left only with the second and forth contributions as follows,

\begin{eqnarray}
\mathcal{L}_{\text{eff}} &=& \frac{g^{2}+g'^{2}}{m_{Z}^{2}} \frac{1}{4}\bar{e}\gamma^{\mu}\gamma^{5}e\left[\left(\frac{1}{4} - \frac{2}{3}\sin^{2}\theta_{W}\right)\bar{u}\gamma_{\mu}u +\right.\nonumber\\
&&\left. \left(-\frac{1}{4} + \frac{1}{3}\sin^{2}\theta_{W}\right)\bar{d}\gamma_{\mu}d \right].
\label{eq:lagrangeffec}
\end{eqnarray}

Hence, we have reviewed that only a vector-axial current to electrons and pure vector interaction with quarks are relevant to APV. This is key to understand how APV can be used to constrain new physics. Notice that this lagrangian describes a neutral current and therefore neutral gauge bosons are natural candidates to induce APV signatures as we investigate further.

As a first step, we give an effective description of new physics on parity violation. We will encode the new physics effects with three parameters namely, $\Lambda$ which represents a new energy scale, and the couplings $f_{V_u}$ and $f_{V_d}$ that quantify how new physics interact with electrons and quarks. In summary, we get,

\begin{eqnarray}
-\mathcal{L}_{\text{eff}}^{\text{PV}} &=& \frac{g^{2}+g'^{2}}{m_{Z}^{2}} \frac{1}{4}\bar{e}\gamma^{\mu}\gamma^{5}e\left[\left(\frac{1}{4} - \frac{2}{3}\sin^{2}\theta_{W}\right)\bar{u}\gamma_{\mu}u \right.\nonumber\\
&& \left. + \left(-\frac{1}{4} + \frac{1}{3}\sin^{2}\theta_{W}\right)\bar{d}\gamma_{\mu}d \right] + \nonumber \\
&&+ \frac{1}{\Lambda^{2}}\bar{e}\gamma^{\mu}\gamma^{5}e\left[f_{Vu}\bar{u}\gamma_{\mu}u + f_{Vd}\bar{d}\gamma_{\mu}d \right]. \label{eq:lagrang}
\end{eqnarray}

Eq.\eqref{eq:lagrang} is a general expression of new physics contributions to parity violation. We could have simply absorbed the coupling constants in the unknown scale of new physics, but we wrote them down explicitly to be as general as possible and to ease the application of our findings to concrete new physics models.\\

Now we have reviewed how one can describe parity violation using effective field theory we will link it to APV.  \\

\section{Atomic Parity Violation}

In order to relate parity violation to APV, we need to define the weak charge of a nucleus which enters into the Hamiltonian of the electron field. The effective Hamiltonian density of the electron field in the vicinity of a nucleus is \cite{Bouchiat:2004sp},

\begin{equation}
\mathcal{H}_{\text{eff}}=  e^{\dagger}(\vec{r}) \gamma_5 e(\vec{r}) \frac{G_F}{2\sqrt{2}}Q_W^{\text{eff}} (Z,N)\delta(\vec{r}),  
\end{equation}where $Q_W^{\text{eff}}$ is the weak-charge of the nucleus which encodes the Standard Model and new physics contributions, $Q_W^{\text{eff}} = Q_{SM}+Q_{NP}$ . The weak charge of a nucleus is defined as $Q_W=( 2f_{V u} + f_{V d})Z +(f_{Vu} + 2 f_{Vd})N$. From this we conclude that the Standard Model and the new physics contributions to the weak charge of a nucleus are,

\begin{eqnarray}
    Q_{SM} &=& \left(\frac{1}{4} - \frac{2}{3}\sin^{2}\theta_{W}\right)(2Z+N) +\nonumber\\
    &&\left(-\frac{1}{4} + \frac{1}{3}\sin^{2}\theta_{W}\right)(Z+2N)\nonumber\\ &&= \frac{1}{4}(Z(1-4\sin^{2}\theta_{W})-N) \equiv \frac{1}{4}Q^{SM}_{W}.
\end{eqnarray}

Consequently the new physics contribution to the weak charge of a nucleus is
\begin{align}
  Q_{\text{NP}} & = \frac{f_{Vu}}{\Lambda^2}\left(2Z+N \right)+\frac{f_{Vd}}{\Lambda^2}\left(Z+2N \right)\nonumber\\ &   = \frac{f_{Vu}\left(2Z+N \right)+f_{Vd}\left(Z+2N \right)}{3(Z+N)\Lambda^2}3(Z+N) \nonumber\\
  &= \frac{3}{\Lambda^2}f_{Vq}^{\text{eff}}(Z+N), 
\end{align}
where the effective coupling is defined as,
\begin{equation}
    f_{Vq}^{\text{eff}} = \frac{f_{Vu}\left(2Z+N \right)+f_{Vd}\left(Z+2N \right)}{3(Z+N)}.
    \label{eq:feff}
\end{equation}

We can then write an effective Hamiltonian in terms of the new effective charge. In the non-relativistic regime, after a Fourier transform the propagator becomes,
\begin{align}
    \frac{1}{m_{Z}^{2}-q^{2}} \rightarrow \frac{\text{e}^{-m_{Z}|\vec{r}|}}{4\pi|\vec{r}|} =\frac{1}{m_{Z}^{2}} \frac{m_{Z}^{2}\text{e}^{-m_{Z}|\vec{r}|}}{4\pi|\vec{r}|}, 
\end{align}
when $m_{Z}^{2}\rightarrow \Lambda^{2}\rightarrow \infty$ we get,
\begin{align}
    \frac{1}{m_{Z}^{2}} \frac{m_{Z}^{2}\text{e}^{-m_{Z}|\vec{r}|}}{4\pi|\vec{r}|} \rightarrow \frac{\delta(\vec{r})}{\Lambda^{2}}.
\end{align}

Then, the effective Hamiltonian for an electron under the weak force of the nucleus which is quantified through the weak charge can be written as,  
\begin{align}
&\left.H_{\text{eff}}^{PV}\right|_{\text{int}}  = -\left.\mathcal{L}_{\text{eff}}^{PV}\right|_{\text{int}}\nonumber\\
& = e^{\dagger}\gamma^{5}e\left[ \frac{g^{2}+g'^{2}}{4m_{Z}^{2}}\frac{1}{4}Q^{\text{SM}}_{W} + \frac{3}{\Lambda^{2}}f_{Vq}^{\text{eff}}(Z+N) \right] \delta(r)\\
&= e^{\dagger}\gamma^{5}e\frac{g^{2}+g'^{2}}{4m_{Z}^{2}}\frac{1}{4}\left[Q^{SM}_{W} + \frac{16m_{Z}^{2}}{g^{2}+g'^{2}} \frac{3}{\Lambda^{2}}f_{Vq}^{\text{eff}}(Z+N) \right]\delta(r)\\
= &e^{\dagger}\gamma^{5}e\frac{G_{F}}{2\sqrt{2}}Q^{\text{eff}}_{W}(Z,N)\delta(r).
\end{align}

The variation of the effective charge is given by,
\begin{align}
    \Delta Q_{W} = Q_{W}^{\text{eff}} - Q_{W}^{\text{SM}} = \frac{2\sqrt{2}}{G_{F}} \frac{3}{\Lambda^{2}}f_{Vq}^{\text{eff}}(Z+N).
    \label{eq:newdif}
\end{align}

The most accurate measurement of APV effect occurs through transitions on the stable isotope $^{133}_{55}Cs$. The prediction for the Standard Model for the weak nuclear charge $Q_{W}$ is, \cite{dzuba2012revisiting},
\begin{equation}
    Q^{\text{th}}_{W} = -73.16(5).
\end{equation}

On the other hand, the measurement yields, 

\begin{equation}
    Q^{\text{ex}}_{W} = -73.16(35),
\end{equation}
and the $90\%$ confidence level difference between the two values is,
\begin{align}
    \Delta Q(Cs) =& \left| Q^{\text{exp}}_{W} - Q^{\text{th}}_{W}\right| <0.6. 
\end{align}

Using equation Eq.\ref{eq:newdif} we can place bounds on new physics effects,
\begin{equation}
     \Delta Q(Cs)=\frac{2\sqrt{2}}{G_{F}} \frac{f_{Vq}^{\text{eff}}}{\Lambda^{2}}3(Z+N)<0.6.
\end{equation}

In case that the new gauge bosons are light (namely masses below order of few MeV) the EFT description we have adopted so far is not valid and one should apply a correction factor $K(Cs)$ \cite{bouchiat2005constraints} to $\Delta Q(Cs)$ in order to properly account the propagator of the new degree of freedom. We will not explicitly consider this case here.
Inserting $G_{F} = 1.1663787\times10^{-5}\, GeV^{-2}$, $Z=55,\, N = 78$, we constrain the ratio effective coupling over the energy scale, 
\begin{equation}
    \frac{f_{Vq}^{\text{eff}}}{\Lambda^{2}}<4.38699\times10^{-9}\, \text{GeV}^{-2}.
\end{equation}

We highlight that this bound is based on the weak charge of the Cesium. The scale $\Lambda$ can be replaced by a heavy mediator mass. Later we will compare this bound with the one stemming from neutrino-nucleus coherent scattering addressed below.

\section{Neutrino-Nucleus Coherent Scattering}

The lagrangian that dictates the APV discussion, Eq.\eqref{eq:lagrang} has the form $\bar{e}\gamma_\mu \gamma_5 e \bar{q}\gamma_\mu q$ with $q=u,d$. From $SU(2)$ invariance, whatever new physics that interacts with electrons should also interact with neutrinos. Therefore, one can use neutrino-nucleus coherent scattering data to constrain the new physics scale where,
\begin{equation}
    -\mathcal{L}_{\text{eff}} = \frac{f_{Vu}}{\Lambda^{2}}(\bar{\nu}\gamma^{\mu}\nu)(\bar{u}\gamma_{\mu}u) + \frac{f_{Vd}}{\Lambda^{2}}(\bar{\nu}\gamma^{\mu}\nu)(\bar{d}\gamma_{\mu}d).
    \label{eqcoherent}
\end{equation}

\begin{figure}[!h]
\centering
\includegraphics[width=\columnwidth]{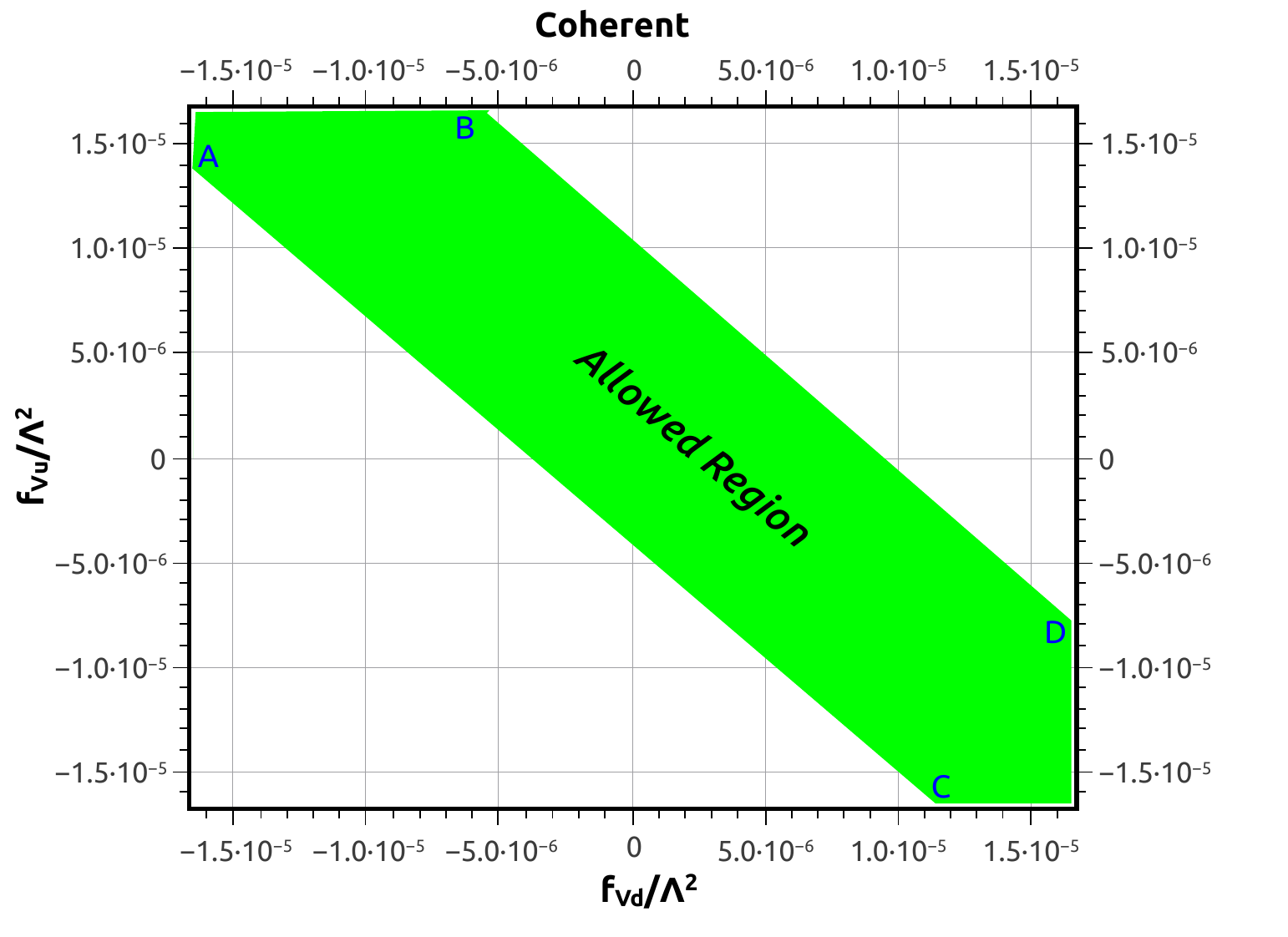}
\caption{Bound on the effective couplings between the electron neutrino and up-quark ($f_{Vu}$) and electron neutrino and down quark ($f_{V_d}$) as defined in Eq.\eqref{eqcoherent}. The green area is the region allowed by the COHERENT data. The benchmark points A, B,C,D shown in the figure will be used to derive limits on $f_{V_q}^{eff}/\Lambda^2$. See the text for details.}
    \label{fig:bounds2}
\end{figure}

COHERENT data is in agreement with the Standard Model prediction. Thus one can use COHERENT data to restrict the presence of new physics effects. Within the effective theory approach one can find the allowed values for the effective couplings $f_{Vu}$ and $f_{Vd}$ defined in Eq.\ref{eqcoherent}. The green area is the allowed region in the $f_{Vu}$ vs $f_{Vd}$ plane \cite{Liao:2017uzy}. In order to compare the sensitivity of neutrino-nucleus coherent scattering and APV, we need to compute $f_{V_q}^{eff}$. The APV observable is related to the coupling between electrons and quarks. Hence, we need to invoke $SU(2)$ invariance and then assume that the effective couplings in Eq.\ref{eq:lagrang} and Eq.\ref{eqcoherent} are similar. Under that assumption, we can pick the pairs $f_{Vu},f_{Vd}$ for the four benchmark points (A,B,C,D) shown in Fig.\ref{fig:bounds2} and then use  Eq.\ref{eq:feff} to find bounds over $f_{V_q}^{eff}/\Lambda^2$.  We conclude from Fig.\ref{fig:bounds2} that the effective coupling should lie between anywhere between AB and CD, which implies in $-2.88\times 10^{-6}<f_{V_q}^{eff}/\Lambda^2< 4\times 10^{-6}$ and $-7.7 \times 10^{-7}<f_{V_q}^{eff}/\Lambda^2< 5.7\times 10^{-6}$, respectively. Taking the most restrictive set of couplings we find that,

\begin{table*}[!t]
    \centering
    \begin{tabular}{|c|c|c|c|}
    \hline
        \hline
     A   &  $f_{Vu}/\Lambda^2=1.47 \times 10^{-5}$ &$f_{Vd}/\Lambda^2=-.162\times 10^{-5}$ & \\
        B & $f_{Vu}/\Lambda^2=1.64 \times 10^{-5}$ & $f_{Vd}/\Lambda^2=-5.38 \times 10^{-6}$ &$-2.88\times 10^{-6}<f_{V_q}^{eff}/\Lambda^2< 4\times 10^{-6}$ \\
         \hline
        C & $f_{Vu}/\Lambda^2=-1.66 \times 10^{-5}$ & $f_{Vd}/\Lambda^2=1.12 \times 10^{-5}$ &\\
       D &$f_{Vu}/\Lambda^2=-8.56 \times 10^{-6}$  & $f_{Vd}/\Lambda^2=1.65 \times 10^{-5}$ &$-7.7 \times 10^{-7}<f_{V_q}^{eff}/\Lambda^2< 5.7\times 10^{-6}$ \\
        \hline
    \end{tabular}
    \caption{Bound on the effective coupling $f_{V_q}^{eff}$ relevant for the APV based on the COHERENT constraint on $f_{V_u}$ and $f_{V_d}$. These bounds assume $SU(2)$ invariance to correlate the effective operators in Eq.\ref{eq:lagrang} and Eq.\ref{eqcoherent}. To find $f_{V_q}^{eff}$ we used Eq.\ref{eq:feff}. Notice that even if we take the most restrictive values for $f_{V_q}^{eff}$ in the table, APV still provides a more stringent bound. }.
    \label{tab:my_label}
\end{table*}

\begin{equation}
    -7.7\times 10^{-7} <\frac{f_{Vq}^{\text{eff}}}{\Lambda^{2}}<4\times 10^{-6}.
\end{equation}

It is exciting to see that COHERENT, a $14$~kg detector, can already place important bounds on new physics. This fact has triggered several new physics sensitivity studies using COHERENT data and other nuclei \cite{deNiverville:2015mwa,Abdullah:2018ykz,Billard:2018jnl,Canas:2018rng,AristizabalSierra:2018eqm,Billard:2018jnl,Farzan:2018gtr,Bischer:2018zbd,Brdar:2018qqj,Denton:2018xmq,Miranda:2019skf,Miranda:2019wdy,AristizabalSierra:2019ufd}. Nevertheless, it is clear that the bound on the effective couplings rising from APV is much more stringent than the one stemming from neutrino-nucleus coherent scattering. It does not undervalue the use of neutrino-nucleus scattering data to constrain new physics because as we previously pointed out they strictly probe different couplings. We emphasize that our conclusion relies on the $SU(2)$ invariance argument. There are upcoming experiments that aim at probing neutrino-nucleus coherent scattering at different energies and precision which will be certainly important to improve the overall sensitivity to new physics \cite{Dent:2016wcr,Dent:2017mpr,Hakenmuller:2019ecb}.

We have parametrized the new physics effect in terms of the effective couplings and the energy scale $\Lambda$. This parametrization is valid in the regime which the new physics scale is much heavier than the typical energy scale involved. However, new physics can also appear as light mediators with masses much below the $Z$ mass. Light mediators alter the Standard Model prediction for $\sin\theta_W$ at low energy. The deviation can be observed using polarized electron scattering as we explore below.

\section{Polarized Electron Scattering}

Atomic Parity Violation is an important probe to test new physics, but if new physics surface at low energy polarized electron scattering becomes an ideal laboratory, specially in the presence of kinetic and mass mixing terms between the $Z$ and $Z^\prime$ gauge boson. Indeed, low energy scattering of polarized electrons on electrons and other targets are very sensitive to parity violation effects at low energy, and consequently to the presence of light mediators. In other words, they are very sensitive to the parity violation effects that are proportional to $1-4\sin^2\theta_W$, and in this way constrain $\sin\theta_W$.  Generally, additional parity violation sources rise from both kinetic and mass $Z-Z^\prime$ mixings. Parametrizing the mass mixing matrix as,\\

\begin{equation}
 M^{2}=\left(\begin{array}{cc}m_{Z}^{2}& -\delta m_{Z}m_{Z'}\\ -\delta m_{Z}m_{Z'}& m_{Z'}^{2}\end{array}\right),
\end{equation}where $0\leq\delta<1$, if the $Z^\prime$ is light compared to Z we can write,

\begin{equation}
M^{2}=\left(\begin{array}{cc} 1& -\delta \frac{m_{Z'}}{m_{Z}}\\ -\delta \frac{m_{Z'}}{m_{Z}}& \frac{m_{Z'}^{2}}{m_{Z}^{2}}\end{array}\right) m_{Z}^{2},
\end{equation}with
\begin{equation}
    \epsilon_Z = \delta \frac{m_{Z'}}{m_{Z}}.
    \label{eqdelta}
\end{equation}

Therefore, the relevant lagrangian that will induce parity violation is found to be,

\begin{equation}
\mathcal{L}=- \left(e\epsilon J_\mu^{em} -\frac{g}{2 \cos\theta_W}\epsilon_Z J_\mu^{NC} \right)Z^\prime,
\end{equation}where $J_\mu^{NC}$ is the Standard Model neutral current. These terms induce weak currents that can accounted for by redefining the $\sin\theta_W$ as \cite{Derman:1979zc,Davoudiasl:2012ag}, 
\begin{equation}
    \sin^{2}\theta_{W}\rightarrow \kappa_{d}\sin^{2}\theta_{W},\qquad \kappa_{d}= 1 - \frac{\epsilon}{\epsilon_{Z}}\delta^{2} \frac{\cos\theta_{W}}{\sin\theta_{W}}f(Q^{2}/m_{Z'}^{2}), \label{eq:replacement}
\end{equation}
where $f(Q^{2}/m_{Z^\prime}^{2})$ is the propagator effect given by  \cite{kumar2013low,armstrong2001qweak,mammei2012moller,souder2012parity},
\begin{equation}
     f(Q^{2}/m_{Z^\prime}^{2}) = \frac{1}{1+Q^{2}/m_{Z^\prime}^{2}}. \label{eq:propeff} 
\end{equation}

Using equations (\ref{eq:replacement}) and (\ref{eq:propeff}) we can write the change of the weak angle due to the mixing between $Z$ and $Z^\prime$ as,
\begin{align}
        \Delta\sin^{2}\theta_{W}\simeq -0.42\epsilon\delta\frac{m_{Z}}{m_{Z^\prime}}f(Q^{2}/m_{Z^\prime}^{2}),
\end{align}and so we can put bounds on the mixing $\epsilon$ given the difference between measurements and prediction of the weak angle,
\begin{align}
      \epsilon^{2}& = \frac{5.66893}{\delta^{2}} (\Delta\sin^{2}\theta_{W})^{2}\left(\frac{m_{Z'}}{m_{Z}}\right)^{2}(1+Q^{2}/m_{Z'}^{2})^{2} \nonumber\\
      &=  \frac{5.66893}{\delta^{2}} (\Delta\sin^{2}\theta_{W})^{2}\left(\frac{m_{Z'}^{2}+Q^{2}}{m_{Z}m_{Z'}}\right)^{2}.
\end{align}

\begin{figure*}[h]
\includegraphics[width=\columnwidth]{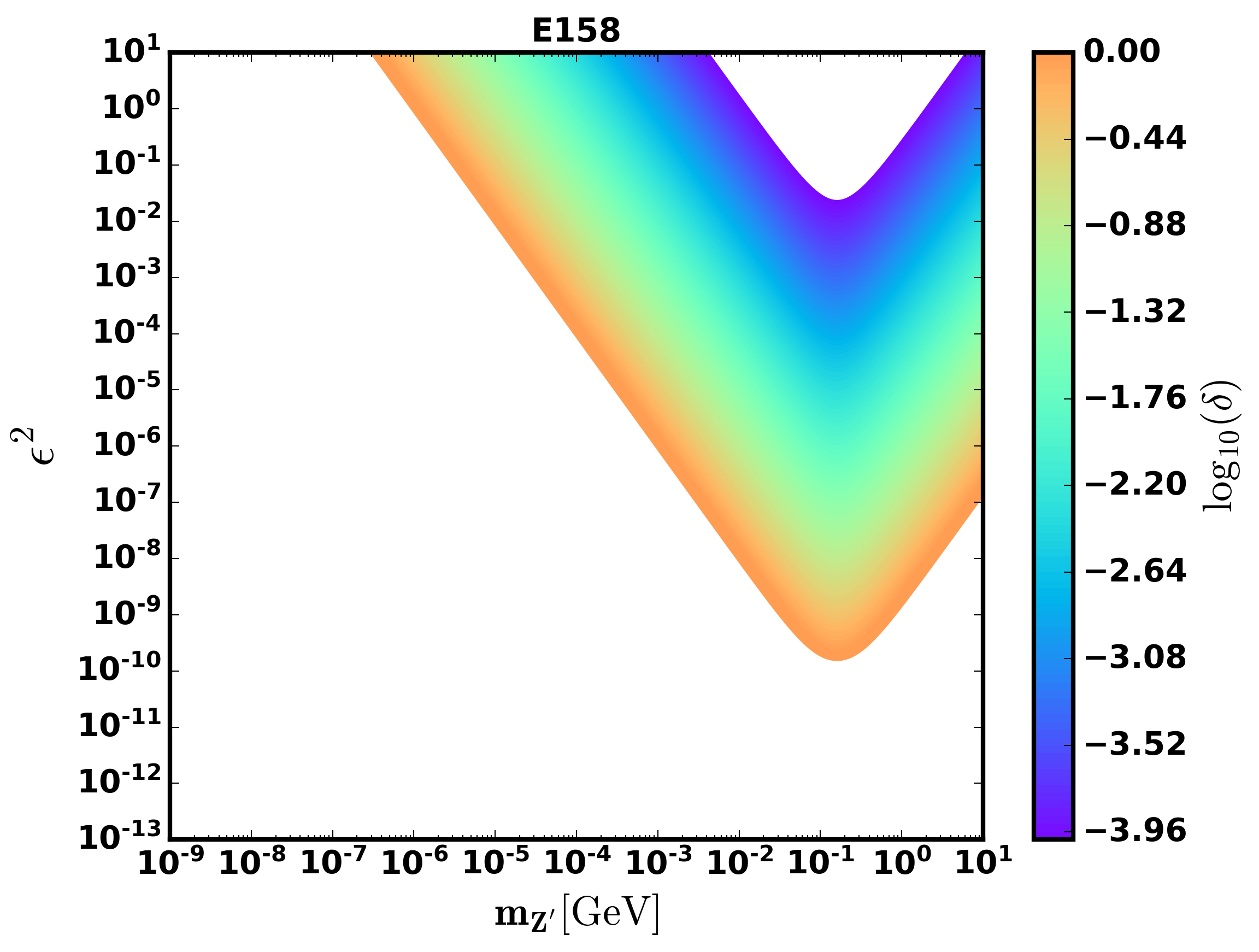}
\includegraphics[width=\columnwidth]{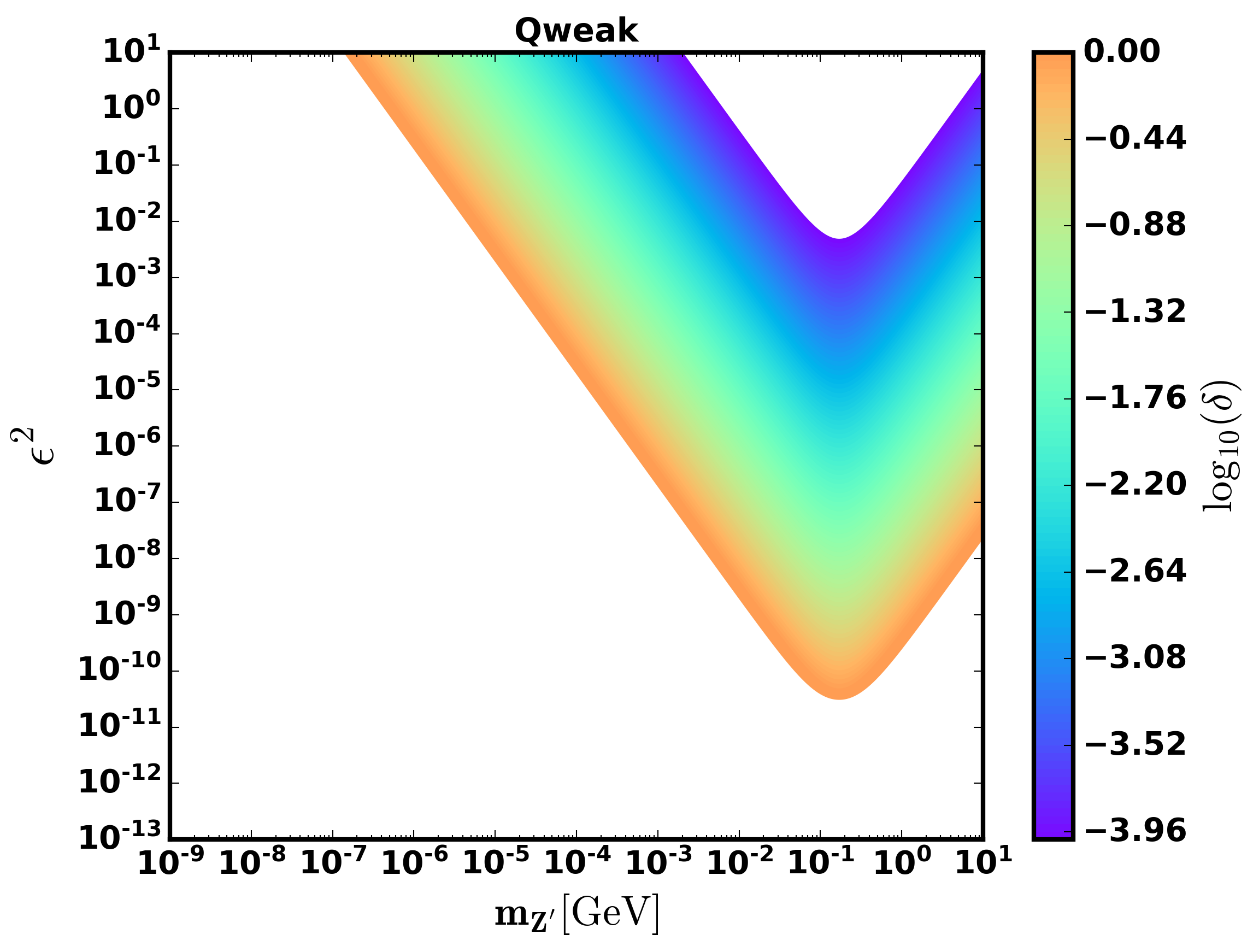}
\includegraphics[width=\columnwidth]{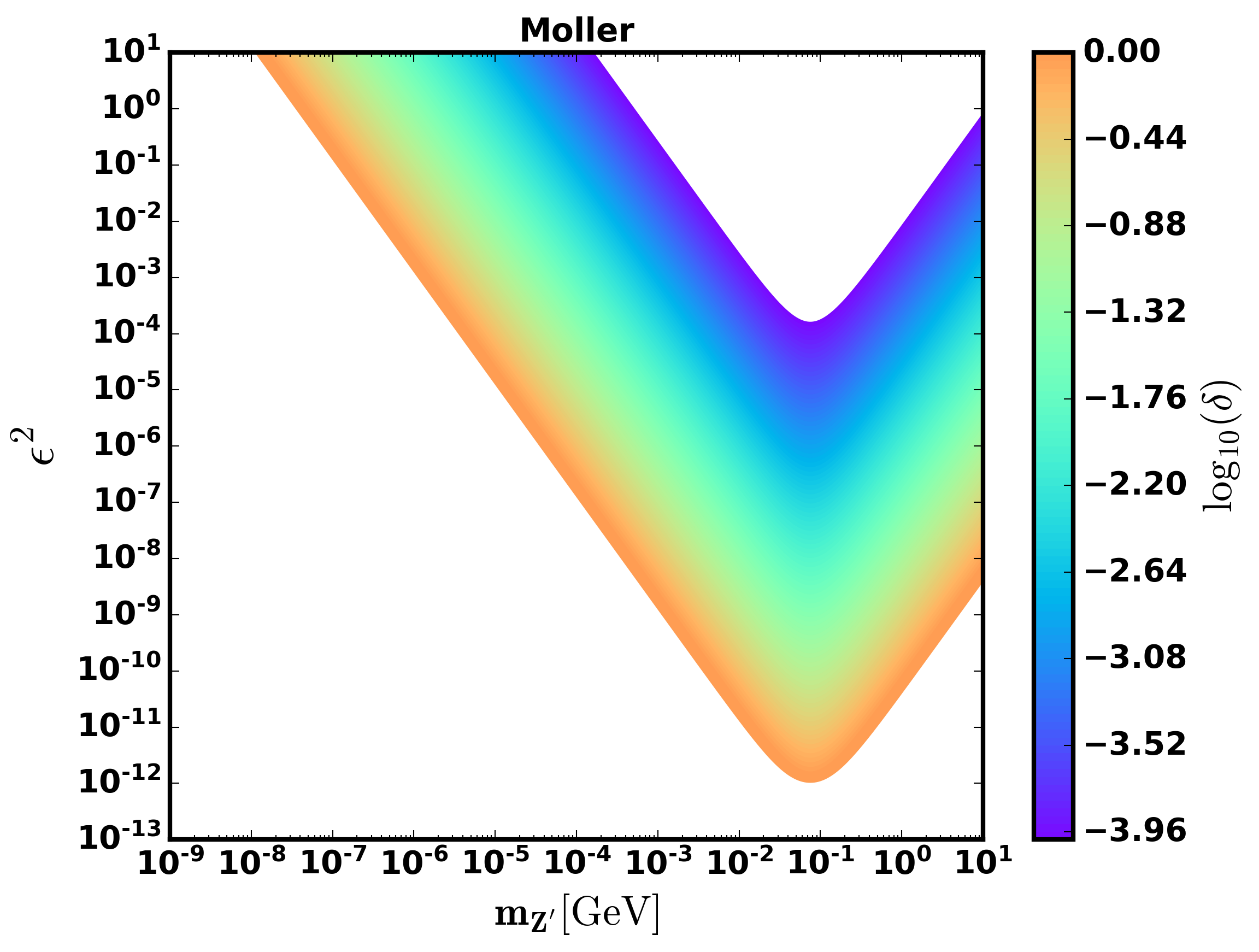}
\includegraphics[width=\columnwidth]{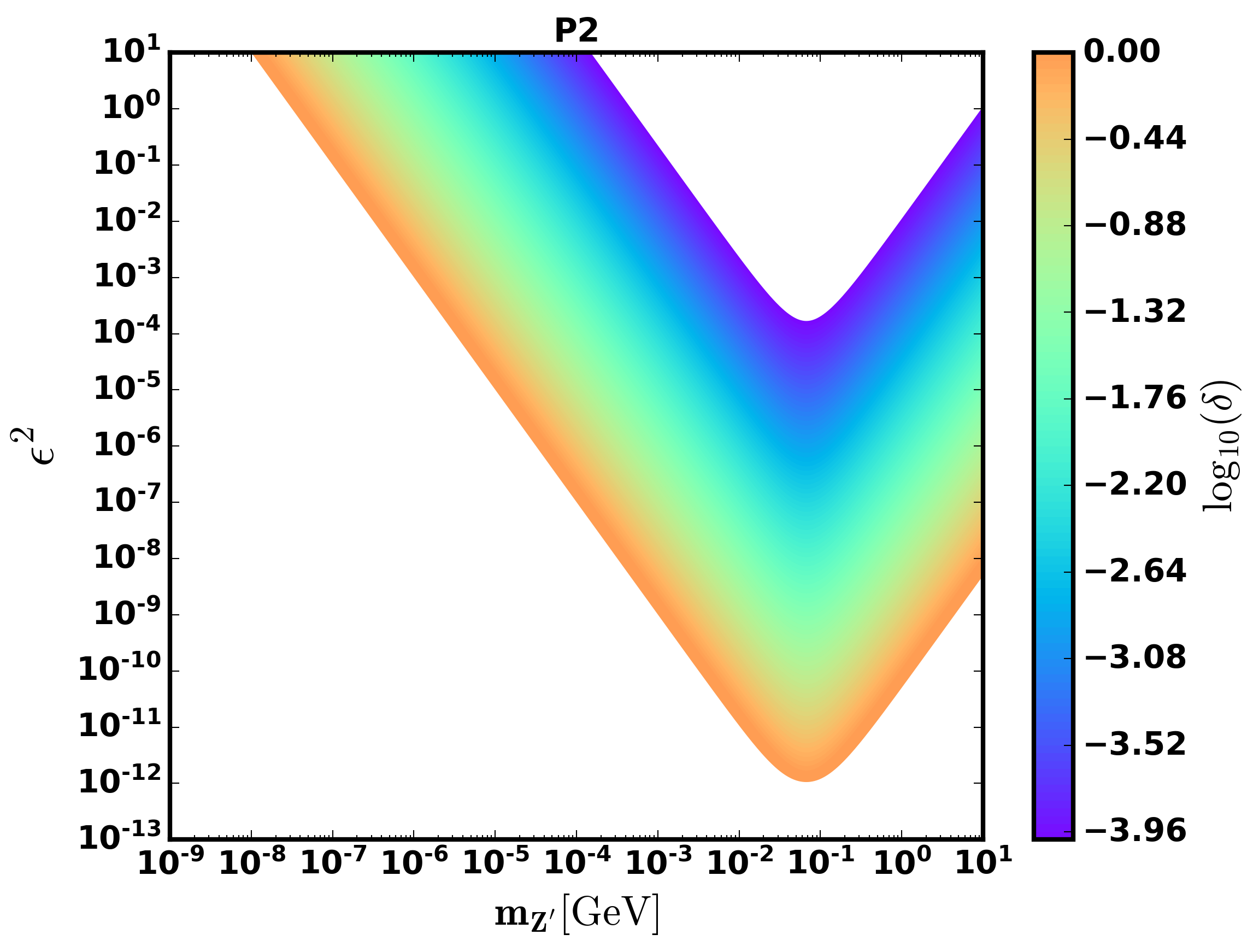}
\includegraphics[width=\columnwidth]{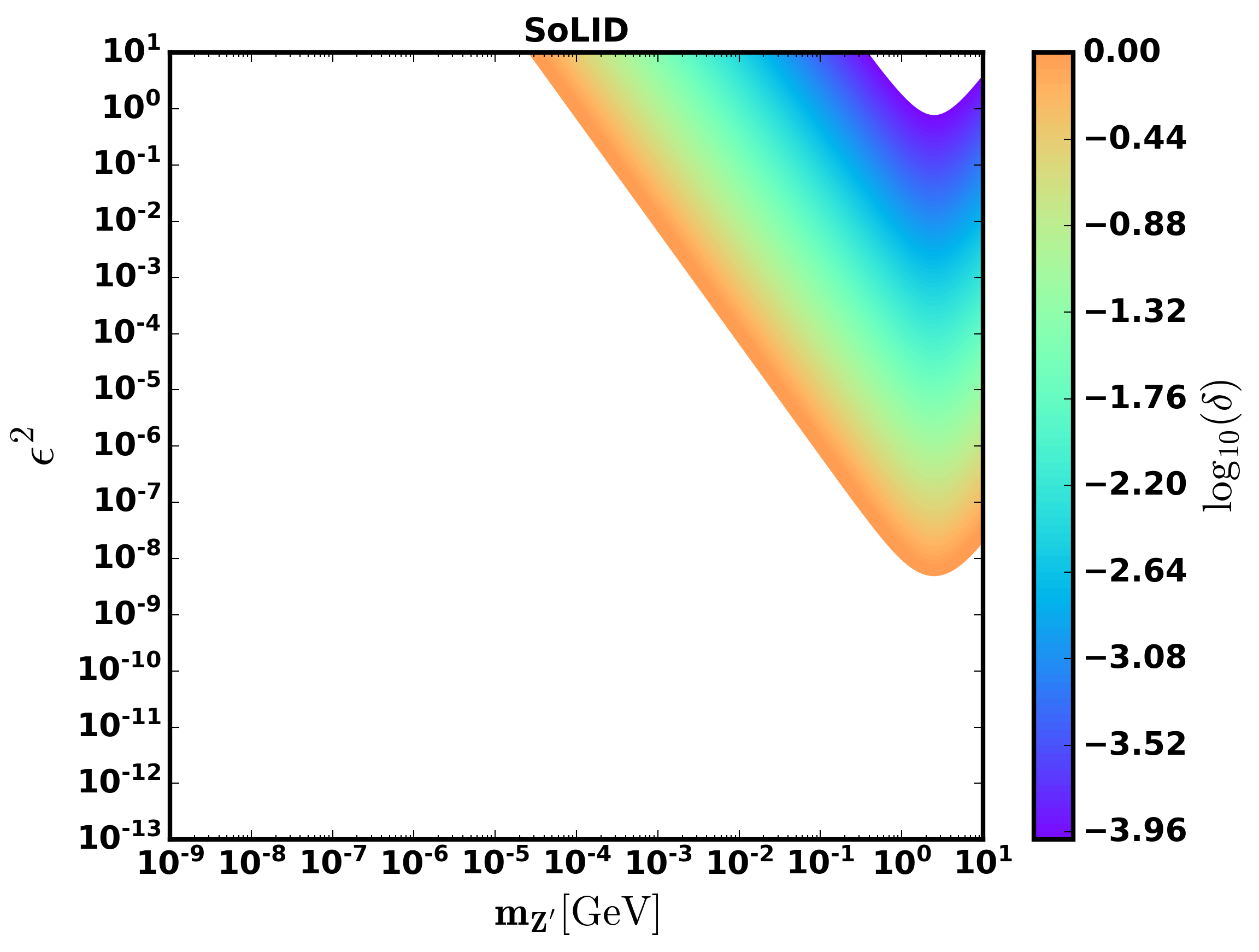}
\caption{Bounds of the experiments E158, Qweak, Moller, P2 and SoLID on the mixing between $Z$ and $Z^\prime$ with respect to the mass of the new light neutral gauge boson following the relations on the table. In each plot the mixing parameter $\delta$ is changing logarithmically from $\delta=10^{-4}$ to $\delta = 1$, so we can also visualize the dependence of $\epsilon$ with this parameter.}
 \label{fig:bounds1}
\end{figure*}

The bounds obtained using the existing and expected precision in the measurement of $\sin^{2}(\theta_{W})$ by some future experiments are written in the {\it Table} \ref{tablePV}. From {\it Table} \ref{tablePV} we notice that the bounds on $\epsilon$ become stronger for large values of $\delta$ which accounts for the mass mixing. We exhibited these bounds for several values of $\delta$ in Fig.\ref{fig:bounds1}. Since the experiments run at different energies they are sensitive to different $Z^\prime$ masses. In particular, SoLID is very sensitive to $Z^\prime$ masses around $1$~GeV. It is remarkable the precision aimed by Moller at JLab planning to measure $\sin^{2}(\theta_{W})$ to $\pm 0.00029$ at $\langle Q\rangle=75$ MeV, followed by the P2 experiment with precision of $\pm 0.00033$ in $\sin^{2}(\theta_{W})$ for $\langle Q\rangle =67$~MeV. Looking either at the {\it Table} \ref{tablePV} or Fig.\ref{fig:bounds1} one can see that if the mass mixing parameter $\delta$ is of the order of $10^{-2}$ precise measurements on $\sin^2\theta_W$ give rise to stringent bounds on $\epsilon$, namely $\epsilon^2 < 10^{-4}$.

\begin{table}[]
\centering
\begin{tabular}{|c|c|c|c|}
	\hline 
	Lab & $\langle Q\rangle$ &  $\sin^{2}\theta_{W}\, (m_{Z})$ & Light Mediator (90\% CL) \\ 
	\hline 
	E158& 160 MeV  & 0.2329(13) &  $\epsilon^{2}<\frac{1.54\times 10^{-5}}{\delta^{2}}\left(\frac{m_{Z'}^{2}+160^{2}}{m_{Z}m_{Z'}}\right)^{2}$ \\ 
	\hline 	Qweak  & 170 MeV & $\pm$ 0.0007 & $\epsilon^{2}<\frac{2.78\times 10^{-6}}{\delta^{2}}\left(\frac{m_{Z'}^{2}+170^{2}}{m_{Z}m_{Z'}}\right)^{2}$ \\ 
	\hline 
	Moller & 75 MeV & $\pm$ 0.00029 & $\epsilon^{2}<\frac{4.77\times 10^{-7}}{\delta^{2}}\left(\frac{m_{Z'}^{2}+75^{2}}{m_{Z}m_{Z'}}\right)^{2}$ \\ 
	\hline 
	P2 & 67 MeV & $\pm$ 0.00033 & $\epsilon^{2}<\frac{6.17\times 10^{-7}}{\delta^{2}}\left(\frac{m_{Z'}^{2}+67^{2}}{m_{Z}m_{Z'}}\right)^{2}$  \\ 
	\hline
	SoLID& 2.5 GeV & $\pm$ 0.0006 & $\epsilon^{2}<\frac{2.04\times 10^{-6}}{\delta^{2}}\left(\frac{m_{Z'}^{2}+2500^{2}}{m_{Z}m_{Z'}}\right)^{2}$ \\
	\hline
\end{tabular} 
\caption{90\% confidence level bounds on the kinetic mixing parameter for light mediators for different experiments that aim at measuring $\sin^2\theta_W$ at low energies. All masses are in $MeV$ units.}
\label{tablePV}
\end{table}

To have an idea of how relevant these limits stemming from $\sin^2\theta_W$ are, we should compare them with those from neutrino-nucleus coherent scattering. The comparison in this case is not so trivial, however, because the bound rising from polarized electron scattering depends on the parameter $\delta$ which is model dependent. Thus, to compare these two methods we need to rely on a specific model.

\section{Models}

In this section will show the relevance of our bounds in anomaly free models that have been proposed in the literature.

\subsection{Dark Z}

The dark $Z$ model proposed in \cite{Davoudiasl:2012ag,Davoudiasl:2012ig,Davoudiasl:2013aya} is an extension of the photon model which includes a free parameter, $\delta$ to account for the existence of a mass mixing term between the $Z$ and $Z^\prime$ gauge bosons as we have done in Eq.\ref{eqdelta}. Such $Z^\prime$ field arises from the presence of a new abelian gauge group. If the Standard Model fermions are uncharged under the $U(1)$ gauge group, the $Z^\prime$ interactions with fermions appear through the presence of kinetic and mass mixing \cite{Campos:2017dgc}. On top of that, the scalar sector is not specified, thus $\delta$ is a free parameter. Looking at {\it Table} \ref{tablePV} we get $\epsilon^2 < 10^{-7}$ for $m_{Z^\prime \sim} 100$~MeV and $\delta \sim 10^{-2}$, using the P2 projected sensitivity. Notice this limit is slightly stronger than the one achieved using {\it BaBar} data \cite{McKenna:2019wim} which is the relevant experiment at this mass range. One can notoriously find stronger bounds from larger values of $\delta$. A study of COHERENT data on neutrino-nucleus coherent scattering yields $\epsilon^2 < 10^{-6}$ \cite{Abdullah:2018ykz}. Therefore, we conclude that our bounds are stronger.  Our bounds are also applicable to the $U(1)_N$ model discussed in \cite{Ko:2012hd,Campos:2017dgc}.

\subsection{Two Higgs Doublet Model}

A UV complete version of the dark photon model has been discussed in the context of a Two Higgs Doublet Models (2HDM) augmented by an Abelian symmetry, $U(1)_X$, \cite{Campos:2017dgc}. In this case, the $Z^\prime$ mass depends on the scalar sector of the model and consequently the parameter $\delta$ is no longer a free parameter. Many $U(1)_X$ extensions where discussed in the context 2HDM, in any case we find $m_{Z^\prime} = g_{X} v \cos\beta^2/\delta$, where $\tan\beta = v_2/v_1$, $v_2 (v_1)$ being the vacuum expectation value of the scalar doublets in the model, and $g_X$ the gauge coupling. The parameter $\delta$ is a function of the $U(1)_X$ charges of the scalars fields and their vacuum expectation values (See Eq.C3 in \cite{Campos:2017dgc}). For most $U(1)_X$ models we find $\delta\sim 10^{-2}$ assuming $\tan\beta \sim 50$. Taking $\delta \simeq 10^{-2}$ we find the bound $\epsilon^2 < 10^{-7}$ for $m_{Z^\prime} \simeq 100$~MeV using the P2 experiment. One can easily recast this limit using the {\it Table} \ref{tablePV}. 

In the heavy mediator regime, we can apply our effective field theory approach taking $f_{V_u} \sim f_{V_d}$ we get $g_{X}^2/m_{Z^\prime}^2 \leq 4.38 \times 10^{-9}$~ GeV. Consequently, $m_{Z^\prime} \geq 1.5 g_{X} \times 10^{4}$~GeV. This bound is applicable under the assumption that vector-axial couplings between the electrons are present as occurs for many models discussed in  \cite{Campos:2017dgc}. Having in mind that LEP bound on vector mediators roughly reads, $m_{Z^\prime} > 7 g_{X}\times 10^3$~GeV \cite{Heeck:2014zfa}, we conclude that APV provides a stronger bound. This limit from LEP was derived for the B-L model where only vectorial interactions are present but it is roughly applicable to other models \cite{Carena:2004xs,Dong:2014wsa}. Anyway, our conclusion stands, APV gives rise to a more restrictive bound on the $Z^\prime$ mass. One may wonder about LHC lower mass bounds on such vector bosons. It has been shown that many of these models predict a large $Z^\prime$ width. This feature weakens LHC sensitivity. Analyzing LHC data it has been found that  
$m_{Z^\prime}> 1-2$~TeV for many models taking $g_X=0.1$ \cite{Camargo:2018klg}, which is again weaker than APV. In summary, APV seems to be the most promising laboratory for such models as far as the $Z^\prime$ mass is concerned.

\subsection{3-3-1 Model}

3-3-1 models are based on the $SU(3)_c \otimes SU(3)_L \otimes U(1)_N$ gauge group \cite{Pisano:1991ee,Foot:1994ym}. They explain the number of replication of fermion generations in the Standard Model and are able to address neutrino masses and dark matter. The presence of the $U(1)_N$ group gives rise to heavy $Z^\prime$ whose mass is set the energy scale at which the 3-3-1 symmetry is broken down to the Standard Model gauge group. The $Z^\prime$ does have vector-axial couplings to electrons and therefore might leave imprints on APV. Although, the $Z^\prime$ couplings to fermions are suppressed, of the order of $10^{-2}$. Using {\it Table V} of \cite{Long:2018fud} where the vector and vector-axial couplings are provided we can compute $f_{V^{eff}_q}$ and consequently find a lower mass bound on the $Z^\prime$ that reads $m_{Z^\prime} > 1.7$~TeV  for the {\it model A} with $\beta=\sqrt{3}$. We point out that the parameter $\beta$ defines the vector and vector-axial couplings in the model according to {\it Table V} in \cite{Long:2018fud}. However, this limit is sub-dominant when compared to existing bounds stemming from dijet and dilepton searches at the LHC which impose $m_{Z^\prime} > 4$~TeV \cite{Coutinho:2013lta,Cao:2016uur,Queiroz:2016gif,Arcadi:2017xbo}. There are other bounds rising from other observables such as from flavor physics but they are not as relevant \cite{Kelso:2014qka,Borges:2016nne,Deppisch:2016jzl,Lindner:2016bgg,Santos:2017jbv,CarcamoHernandez:2018iel}. We highlight that there are possible extensions of this model via the inclusion of right-handed neutrinos which can weaken the LHC bounds by decreasing the $Z^\prime$ branching ratio into charged leptons and quarks \cite{Freitas:2018vnt}. In that case our bounds could become competitive.

\subsection{Light $Z^\prime$ Models}

It has been recently proposed a model which successfully accommodates neutrino masses within the type II seesaw while hosting a light $Z^\prime$ gauge boson \cite{Camargo:2018uzw}. The mass of the $Z^\prime$ comes from the vacuum expectation value of the scalar doublets and therefore the $Z^\prime$ must light. Again the mass mixing parameter $\delta$ depends on the scalar spectrum which is set by anomaly cancellation requirements. For the $U(1)_{Y^\prime}$ presented in {\it Table 1} of  \cite{Camargo:2018uzw} we get $\delta \sim 10^{-1}$. For such a large value of delta we find $\epsilon^2 < 10^{-10}$ for $m_{Z^\prime} \simeq 100$~MeV using the P2 experiment, and $\epsilon^2 < 10^{-7}$ for $m_{Z^\prime} \simeq 1$~GeV using SoLID projection. These bounds are much stronger than those derived using  The Heavy Photon (HPS) Search Experiment and Belle projections shown in \cite{Camargo:2018uzw}. 

Models based on the $L_\mu-L_\tau$ gauge symmetry have recently brought a lot of attention due to some flavor anomaly \cite{Chen:2017usq,Baek:2017sew,Arcadi:2018tly,Foldenauer:2018zrz,Ko:2019tts,Escudero:2019gzq,Biswas:2019twf}. The $Z^\prime$ boson can be quite light and has no interactions with quarks at tree-level. At loop level, one could nevertheless generate the neutrino-nucleus coherent scattering and the parity-violating observables discussed here. Albeit, there are already stringent bounds rising from neutrino-trident production and meson mixings \cite{Altmannshofer:2014pba,Altmannshofer:2016jzy}, making our assessment of 1-loop induced parity violation effects not relevant, in agreement with \cite{Abdullah:2018ykz}.

\section{Conclusions}

We have reviewed the theoretical aspect of parity violation and put it in context with other relevant observables. We treated Atomic Parity Violation using effective field theory and showed how one can constrain new physics via precise measurements of the Cesium weak charge. Moreover, we have discussed neutrino-nucleus coherent scattering and shown that Atomic Parity Violation leads to a more restrictive bound on the new physics scale under the assumption that the new physics particle couples to electrons and neutrinos with similar strength. This conclusion is also valid to heavy vector mediators with masses at the TeV scale, for instance. Shifting the discussion to light mediators we have parametrized new physics effects in polarized electron scatterings in terms of the $\sin\theta_W$ and explored the sensitivity of new measurements on $\sin\theta_W$ to derive bounds on the kinetic mixing between the Z and $Z^\prime$ gauge bosons as a function of the $Z^\prime$ mass.  Lastly, we applied our constraints to models previously proposed in the literature and showed that our findings constitute, in some cases, the strongest limits on the kinetic mixing parameter, highlighting the importance of our reasoning. 

\section*{Acknowledgements}
The authors thank Omar Miranda, Alex Dias, Carlos Pires, Diego Cogollo, and Paulo Rodrigues for feedback on the manuscript.  FSQ acknowledges support from CNPq
grants 303817/2018-6 and 421952/2018-0, UFRN, MEC and
ICTP-SAIFR FAPESP grant 2016/01343-7.

\bibliography{ref}
\end{document}